\documentclass[doublecol]{epl2}

\title{Ordering Kinetics of Canted and Uniform States in \\Nematic Liquid Crystals}

\author{Nishant Birdi\inst{1} \and Varsha Banerjee\inst{1,2} \and Sanjay Puri\inst{1,3}}
\shortauthor{Nishant Birdi \etal}

\institute
{                    
  \inst{1} School of Interdisciplinary Research, Indian Institute of Technology, Hauz Khas, New Delhi -- 110016, India\\
  \inst{2} Department of Physics, Indian Institute of Technology, Hauz Khas, New Delhi -- 110016, India\\
  \inst{3} School of Physical Sciences, Jawaharlal Nehru University, New Delhi -- 110067, India
}

\pacs{nn.mm.xx}{First pacs description}
\pacs{nn.mm.xx}{Second pacs description}
\pacs{nn.mm.xx}{Third pacs description}

\abstract
{We undertake a comprehensive Monte Carlo (MC) study of the ordering kinetics in nematic liquid crystals (NLCs) in 3-dimensions $(d=3)$ by performing deep quenches from the isotropic $(T>T_c)$ to the nematic $(T<T_c)$ phase. The inter-molecular potential between the nematogens, represented by continuous $O(3)$ spins with inversion symmetry, is accurately mimicked by the {\it generalised Lebwohl Lasher} (GLL) model. It incorporates second and fourth order Legendre interactions, and their relative interaction strength is $\lambda$. For $\lambda <-0.3$, we observe {\it canted} morphologies with a $\lambda$-dependent angle-of-tilt between the neighbouring rod-like molecules. For $\lambda \geq-0.3$, the molecules align to yield {\it uniform} states. The coarsening morphologies obey {\it generalized dynamical scaling} in the two regimes, but the scaling function is not robust with respect to $\lambda$. The structure factor tail in the canted regime follows the {\it Porod law}: $S(k,t)\sim k^{-4}$, implying that the coarsening dynamics is due to the annihilation of interfacial defects. This is unexpected, as the GLL model is characterised by a continuous order parameter. The uniform regime on the other hand, exhibits the expected {\it generalized Porod decay}: $S(k,t)\sim k^{-5}$, characteristic of scattering from {\it string defects}. Finally, the domain growth obeys the {\it Lifshitz-Allen-Cahn law}: $L(t)\sim t^{1/2}$ for all values of $\lambda$. Our results for the novel {\it canted} regime are relevant for a large class of systems with orientational ordering, e.g. active matter, membranes, LC elastomers, etc. We hope that our work triggers-off stimulating investigations in them.}

\begin{document}

\maketitle

Liquid crystals (LCs) are a state of matter that is intermediate between conventional solids and liquids \cite{Stephen_1974,deGennes_1995,Priestly_2012,Andrienko_2018}. They are anisotropic in nature, and combine the fluidity of liquids with the long-range order of solids. Chemically, LCs are organic compounds consisting of carbon, hydrogen, oxygen and nitrogen. Based on the structural arrangement of their constituent elements, the LC molecules are broadly classified as calamitic (rod shaped), discotic (disc shaped) and bent-core (banana shaped). Calamitic LCs are the simplest in this family, with tremendous applications in electro-optic devices - the most significant being liquid crystal displays (LCDs) \cite{Chen_2018}. At high temperatures, they exhibit an {\it isotropic} (I) phase characterized by randomly oriented rods (or nematogens), having translational as well as rotational symmetry. At low temperatures, the nematogens align statistically parallel in an arbitrary direction to yield the {\it nematic} (N) phase. It is characterized by long-range orientational order which extends over thousands of molecules. If the N phase is {\it uniaxial}, it is described by a sign-invariant unit vector known as the nematic {\it director} $\textbf{n}$. 

LCs are experimentally accessible systems possessing continuous symmetry, and allow the study of topological defects which are of great interest to the scientific community \cite{Chuang_1991,Pargellis_1991,Lavrentovich_2001,Wang_2016,Kim_2018,Sandford_2020}. A number of approaches, ranging from continuum free energy models to off-lattice and on-lattice models have come forth to understand the emergent phases and their optical, electrical and magnetic responses \cite{Andrienko_2018}. The quest for a unified theoretical framework persists for this fundamental and technologically rich system, to not only understand experimental observations in conventional LCs, but to also include novel LC-based materials such as the biaxial bentcores \cite{Reddy_2006} and the ferromagnetic nematics \cite{Mertelj_2013}. 

One of the early theoretical frameworks to understand the I-N phase transition was provided by Maier and Saupe (MS) \cite{Maier_1958,Saupe_1968}. They used an approximate molecular field theory to obtain the pseudo-potential between two rod-like nematogens. The MS theory did predict the I-N transition to be first order, but the quantitative evaluations for the orientational order parameter were far from accurate. Consequently, Humphries {\textit et al.} extended the MS theory by expressing the pseudo-potential in terms of a complete set of Legendre polynomials $\ P_l(z)$  \cite{Humphries_1972}.  Due to the inversion symmetry of the rod-like molecules, this expansion contains only the even order polynomials ($P_2, \ P_4,$ etc.), and the coefficients are found to decrease rapidly with increasing $l$ \cite{Sweet_1967}.  The MS theory is recovered if the summation in the expansion is restricted to only $l=2$. The simplest three dimensional ($d=3$) lattice version of the MS theory is the Lebwohl-Lasher (LL) model \cite{Lebwohl_1972}. Each lattice site $i$ has a rod-like molecule with orientation of the long axis defined by a three-component ($n=3$) unit vector $\boldsymbol{S}_i=\left(\sin\vartheta_i\cos\varphi_i, \sin\vartheta_i\sin\varphi_i, \cos\vartheta_i\right)$ with inclination $\vartheta_i \in [0,\pi]$ and azimuth $\varphi_i \in [0,2\pi)$. The Hamiltonian for the LL model is given by:
\begin{equation}
\label{equation_1}
H = -J\sum_{\langle ij \rangle}P_2\left(\cos\theta_{ij}\right),
\end{equation}
where $J$ is the strength of the nearest neighbour (nn) interactions, $\theta_{ij}$ is the angle between the nn nematic rods ($\cos\theta_{ij}=\boldsymbol{S}_i\cdot\boldsymbol{S}_j$) and $P_2(z)=\left(3z^2-1\right)/2$. In subsequent discussions, we set $J=1$ for convenience.  The orientational order parameter is given by:
\begin{equation}
\label{equation_2}
\mathcal{S} = \langle P_2\left(\cos\theta_{i}\right)\rangle = \left\langle \frac{3\cos^2\theta_{i}-1}{2}\right\rangle.
\end{equation}
In the above expression, $\cos\theta_i = \boldsymbol{S}_i\cdot \boldsymbol{n}$ and the angular brackets $\langle\cdot\cdot\cdot\rangle$ imply an ensemble average. The isotropic phase corresponds to $\mathcal{S}=0$ while a fully aligned nematic phase has $\mathcal{S}=1$. A defect corresponds to a region of low order or $\mathcal{S}\simeq 0$.

{\color{black} The LL model has certain disadvantages inherent in a lattice model. For example, the translational motion of molecules is eliminated, so that one cannot discriminate between a liquid crystal and a crystal. It is unclear whether the structures seen in the LL model or its generalizations would survive if molecular motion were permitted. Further, diffusive motion and hydrodynamics is also absent, along with the corresponding conservation laws. Nevertheless, it has been widely used in the literature to understand the nematic mesophase because of it's simplicity, analytical tractability and computational ease. The LL model exhibits a {\it weak} first order I-N transition, and inherits the quantitative discrepancies of the MS theory \cite{Zhang_1992}.} 

The shortcomings of the LL model have been overcome by the inclusion of higher order long-range and short-range interactions. Scattering experiments indicate that only the terms upto $P_4$ in the expansion of the pseudo-potential are statistically significant \cite{Hamley_1996}. Consequently, a reasonable approximation representing the interaction between neighbouring nematogens is the {generalized} Lebwohl-Lasher (GLL) model, described by the Hamiltonian \cite{Zhang_1993_MolPhys,Chiccoli_1997}:
\begin{equation}
\label{equation_3}
{\color{black}H = \sum_{\langle ij \rangle}-\big[P_2\left(\cos\theta_{ij}\right)+\lambda P_4\left(\cos\theta_{ij}\right)\big] \equiv \sum_{\langle ij \rangle}E(\cos\theta_{ij})},
\end{equation}
where $\lambda$ is the relative strength of the $P_4$ interactions with respect to the $P_2$ interactions. It is a measure of anisotropy in the system and can be positive or negative. The Legendre polynomial $P_4(z)=\left(35z^4-30z^2+3\right)/8$. {\color{black} There are very few studies of the GLL model, and these are limited to investigations of the effect of the $P_4$ term on the isotropic-ordered phase transition \cite{Fuller_1985,Zhang_1993_MolPhys,Chiccoli_1997,Pelcovits_2001}. With increasing strength of the $P_4$ term, the transition temperature $T_c(\lambda)$ increases and the transition becomes first-order. Further, the inclusion of the $P_4$ term improves agreement between experimental observations and numerical results.}
 
In this letter, we perform MC simulations with non-conserved order parameter dynamics to study the kinetics of phase transition or {\it coarsening} in the $d=3$ NLCs. We study this important non-equilibrium phenomenon using the GLL model, motivated by it's success of accurately describing the equilibrium properties. The phenomenon of coarsening is initiated when a system is rendered thermodynamically unstable following a quench of an intensive thermodynamic state variable such as temperature or pressure. The initial symmetry of the disordered phase gets spontaneously broken at the onset of order. A slow non-equilibrium evolution begins via the emergence of correlated regions of competing equilibrium phases or {\it domains} separated by interfaces or {\it defects}. The domains grow in size via annihilation of defects \cite{Bray_2002,Puri_2004,Puri_2009}. This far-from-equilibrium progressive dynamics is usually termed as {\it domain growth} or {\it coarsening}. The nature of the ordering dynamics depends upon several factors such as the symmetry of the order parameter, conservation laws, relevance of hydrodynamics, etc. The growth laws also provide important insights about the free-energy landscape and relaxation time-scales in the system. 

The main observations from our study are as follows:\\
(a) For $\lambda<-0.3$, the system exhibits {\it canted} ground states (GS) with a $\lambda$-dependent angle of tilt between the nn nematic rods. For $\lambda \geq -0.3$ on the other hand, {\it uniform} GS with a parallel alignment of nn rods are observed.  \\   
(b) The domain growth exhibits dynamical scaling in the two regimes. The ordering dynamics in each regime is characterized by a unique length scale, and the morphology of the domains does not change with time apart from a scale factor. The scaling functions in the canted and uniform regimes are distinct, indicating that the morphology is {\it not robust} with respect to $\lambda$. \\
(c) The structure factor for the canted domain morphologies exhibits an unexpected {\it Porod decay}, $S(k) \sim k^{-(d+1)}$, characteristic of scattering from sharp interfaces. The uniform morphologies on the other hand, exhibit the expected {\it generalized Porod decay}, $S(k) \sim k^{-(d+2)}$, due to scattering from string defects typical of the continuous order parameter $O(n)$ models with inversion symmetry. \\
(d) In both regimes, the domain growth obeys the {\it Lifshitz-Allen-Cahn} (LAC) law $L(t)\sim t^{1/2}$, characteristic of systems with non-conserved order parameter kinetics. The growth is faster with increasing values of $\lambda$. \\
The GLL model offers a simple framework to study geometric incompatibilities arising in orientational ordering. Such systems are ubiquitous, and our results therefore are relevant for the large class of complex systems with orientational order.

Coarsening in $d=2$ NLCs is well studied, using coarse-grained free energy models \cite{Zapotocky_1995,Yeomans_2001,Bhattacharjee_2008} as well as discrete models such as the LL \cite{Blundell_1992,Toyoki_1993,Dutta_2005,Singh_2012,Singh_2014} and the GLL\cite{Singh_2013}. These studies predict the LAC law for domain growth, with logarithmic corrections. The primary defects in $d=2$ are vortices and anti-vortices. There are limited studies in $d=3$ using coarse-grained free energy models \cite{Bray_1993,Wickam_1997}, lattice models such as the LL \cite{Blundell_1992,Zhang_1993_PRE,Bradac_2011} and the off-lattice Gay Berne (GB) model \cite{Billeter_1999}. The domain growth obeys the LAC law, and strings and hedgehogs are the dominant {\color{black}defects}.  

\begin{figure*}[htb]
	\centering	
	\onefigure[scale=0.25]{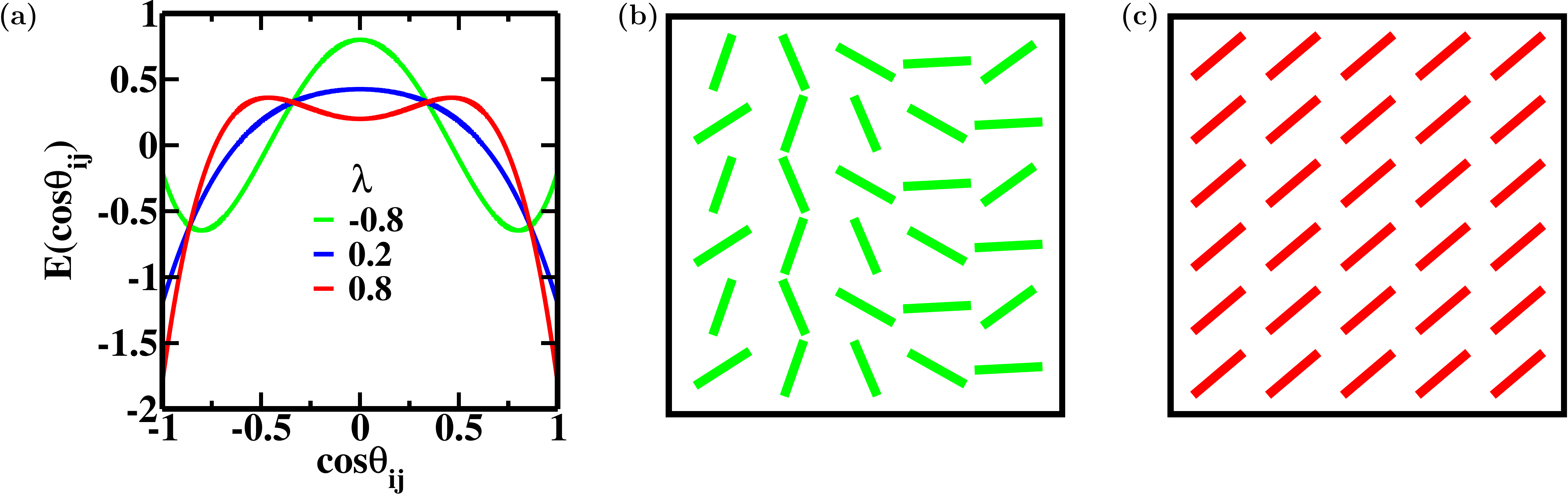}
	\caption{(a) Plot of the energies of the nn pairs in the GLL model, $E(\cos\theta_{ij})$ vs. $\cos\theta_{ij}$, for different values of $\lambda$. Schematic for the (b) canted and (c) uniform ground states. In (b), all the nn rods have an angle of $\simeq36.7^\circ$ ($\lambda=-0.8$) between them and the rods show a rotating or a zig-zag arrangement.}
	\label{figure_1}
\end{figure*}

Let us first examine the GS $(T=0)$ of Eq.~(\ref{equation_3}) as a function of $\lambda$. Fig.~\ref{figure_1}(a) shows the variation of $E(\cos\theta_{ij})$ vs. $\cos\theta_{ij}$ for three different values of $\lambda=-0.8,0.2,0.8$. Using the standard energy minimization procedures, it is easy to see that this model has two distinct classes of GS: (i) For $\lambda < -0.3$, the global minimum of the energy results for $\theta_{ij}=\theta_c=\cos^{-1}\left[\left(15\lambda-6\right)/\left(35\lambda\right)\right]^{1/2}$ (or $\pi-\theta_c$). The variation of $\theta_c$ for different values of $\lambda$ is specified in Table.~\ref{table_1}. Thus, in the GS, the nn nematic rods are aligned at an angle $\theta_{c}$ which is unique to each value of $\lambda$. Note that the subsequent nn rods can form rotating or zig-zag arrangements, see the schematic in Fig.~\ref{figure_1}(b) for $\lambda = -0.8$ ($\theta_c=36.69^\circ$). We refer to these as the {\it canted} GS. To the best of our knowledge, they have not been investigated in the literature. (ii) For $\lambda\geq-0.3$, the global minimum of the energy is observed at $\theta_{ij} = \theta_{u}=0$ (or $\pi$), as seen in the schematic in Fig.~\ref{figure_1}(c). We refer to it as {\it uniform} GS. For $\lambda>0.4$, there is a local minimum of the energy at $\theta_{ij} = \pi/2$ corresponding to a perpendicular alignment of rods which is metastable. It has consequences on the early stage of domain growth, as will be seen in due course. {\color{black}It should be noted that any configuration where the angle between {\it nn} rods is $\theta_c$ or $\theta_u$ as the case may be, is a GS. There are infinite GS corresponding to any value of $\lambda$. }

\begin{table}[h]
\caption{Ground state minimum in the canted regime, $\theta_c$ (in degrees) for different values of $\lambda$.} \centering
\vspace{0.5cm}
\begin{tabular}{|p{0.5cm} |p{7.0cm}|}  
\hline
\vspace*{0.001cm}
$\lambda$ &\vspace*{0.001cm}{ -1.0 \hspace{0.25cm} -0.9 \hspace{0.25cm} -0.8 \hspace{0.25cm} -0.7 \hspace{0.25cm} -0.6 \hspace{0.25cm} -0.5 \hspace{0.25cm} -0.4}\\
\hline
\vspace*{0.001cm}
$\theta_c$ &\vspace*{0.001cm}{39.23 \hspace{0.015cm} 38.11 \hspace{0.015cm} 36.69 \hspace{0.015cm} 34.84 \hspace{0.015cm} 32.31 \hspace{0.015cm} 28.56 \hspace{0.015cm} 22.21}\\
\hline
\end{tabular}
\label{table_1}
\end{table}

We now proceed to study the domain growth kinetics in the $d=3$ GLL model after a quench from the disordered phase $(T>T_c)$ to the ordered phase $(T<T_c)$. These quenches have been guided by the equilibrium phase diagram obtained by Chiccoli {\textit et al.} using cluster MC simulations \cite{Chiccoli_1997}. {\color{black} We consider simple cubic lattices of size $N=N_L^3=(256)^3$ with periodic boundary conditions, and typically quench the system to $T=0.5~T_c(\lambda)$. These deep quenches ensure that the non-equilibrium evolution is not influenced by the properties of the phase transition.} The initial configuration is chosen to be of randomly oriented nematic rods, corresponding to the $T=\infty$ phase. {\color{black} The non-equilibrium evolution is studied via MC simulations. They provide an efficient route to the equilibrium state as compared to the alternative Molecular Dynamics (MD) simulations  \cite{Care_2005}. A randomly chosen nematogen $\boldsymbol{S}_i$ is rotated by a small amount and the new orientation is accepted with standard Metropolis acceptance rates} \cite{Landau_2005}. Each data set has been averaged over 20 different initial conditions. Error bars are not indicated in the figures as the standard errors in the data are smaller than the symbol sizes.

\begin{figure}[h]
	\onefigure[scale=0.31]{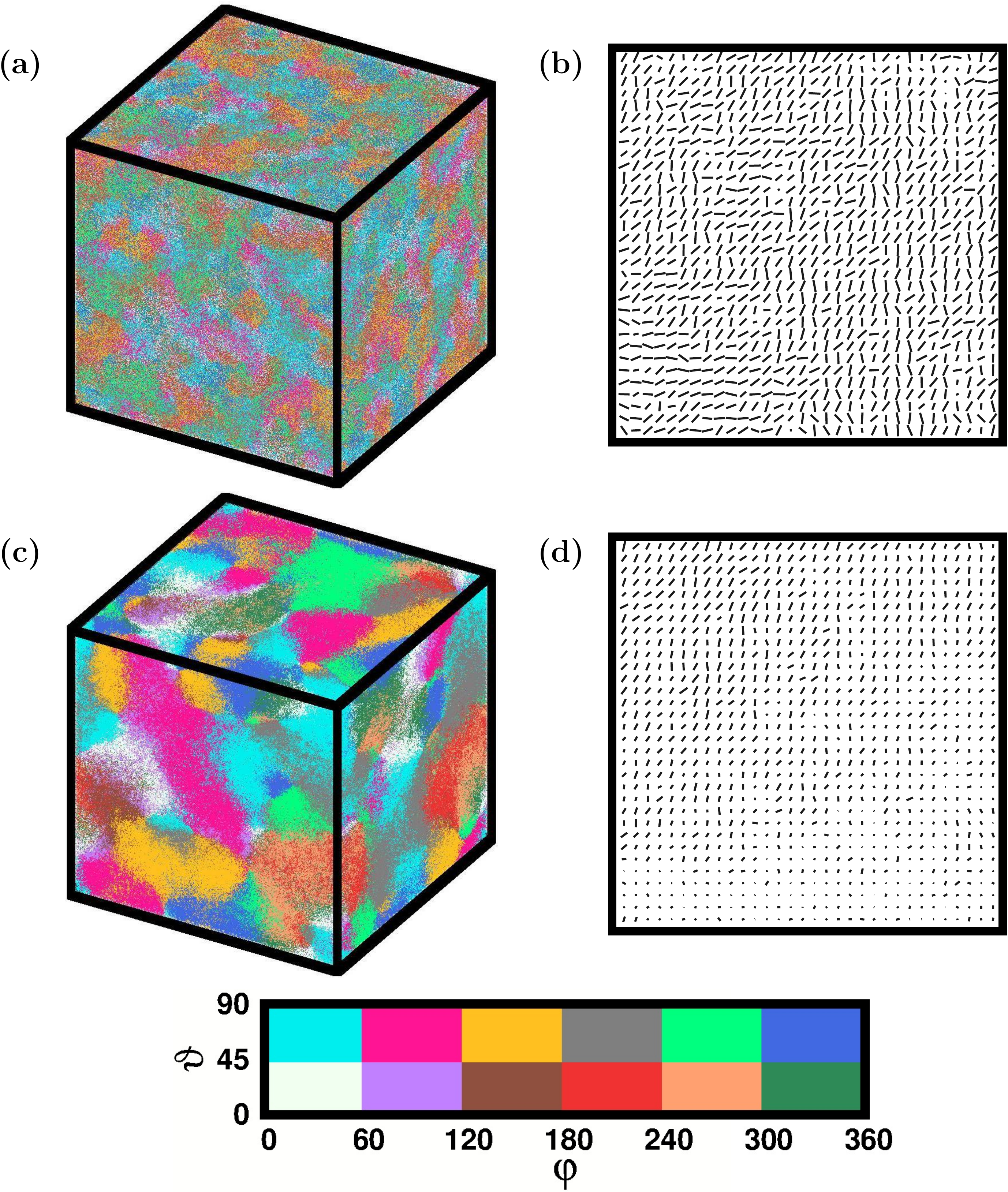}
	\caption{Domain evolution snapshots of the GLL model for (a) $\lambda=-0.8$ and (c) $\lambda=0.8$ at $t=10^5$. The lattice size is $256^3$. [(b) and (d)] Orientations of the nematic rods in the horizontal ($xy$) slices (at $z=N_L/2=128$) of the evolution snapshots  shown in (a) and (c). For a better and clear view, we show only a $32^2$ corner of the entire $256^2$ slice. These rods depict the formation of canted and uniform states, for the respective $\lambda$ values. The key explaining the angular orientations ($\vartheta,\varphi$ in degrees) of the nematic rods in the evolution snapshots shown in (a) and (c) is given at the bottom of the figure.}
	\label{figure_2}
\end{figure}

In Fig.~\ref{figure_2}, we show prototypical domain evolution snapshots (a) in the {\it canted} regime for $\lambda=-0.8$ and (b) the {\it uniform} regime for $\lambda=0.8$ at $t$ (in MCS) $=10^5$. The different colors correspond to the orientation $(\vartheta,\varphi)$ of the nematic rods and are specified in the key. (The range of $\vartheta$ is halved due to the inversion symmetry of the nematogens.) 
For $\lambda = -0.8$, the relative orientation between the nn nematic rods in the GS is $\theta_c\simeq36.7^\circ$ (or $143.3^\circ)$, see Table I. Hence the domains in Fig.~\ref{figure_2}(a) are not monochromatic because of the canted arrangements of the nematogens. For clarity, we explicitly show the orientation of the nematic rods in a horizontal ($xy$) slice taken at $z=N_L/2=128$ in Fig.~\ref{figure_2}(b).  (Only a $32^2$ corner of the entire $256^2$ slice is shown for a distinct view). The morphology in Fig.~\ref{figure_2}(c) on the other hand, is characterized by monochromatic (similar orientation) domains as the corresponding GS has $\theta_{u}=0$ (or $\pi$). The corresponding horizontal slice in Fig.~\ref{figure_2}(d) confirms the emergence of correlated regions of parallel rods characteristic of the uniform regime. {\color{black}We mention here that canted states have been observed in assemblies of rod-shaped particles with long-range interactions. For instance, MD simulations of nematogens with Gay Berne interactions showed ``shape frustrations'' and ``rippled'' (or canted) structures in the smectic phase in concurrence with experimental observations  \cite{Parker_1997,Withers_2002}. It is interesting to note that the canted states are preserved even in these off-lattice descriptions incorporating hydrodynamics.}

\begin{figure*}[htb]
	\onefigure[scale=0.25]{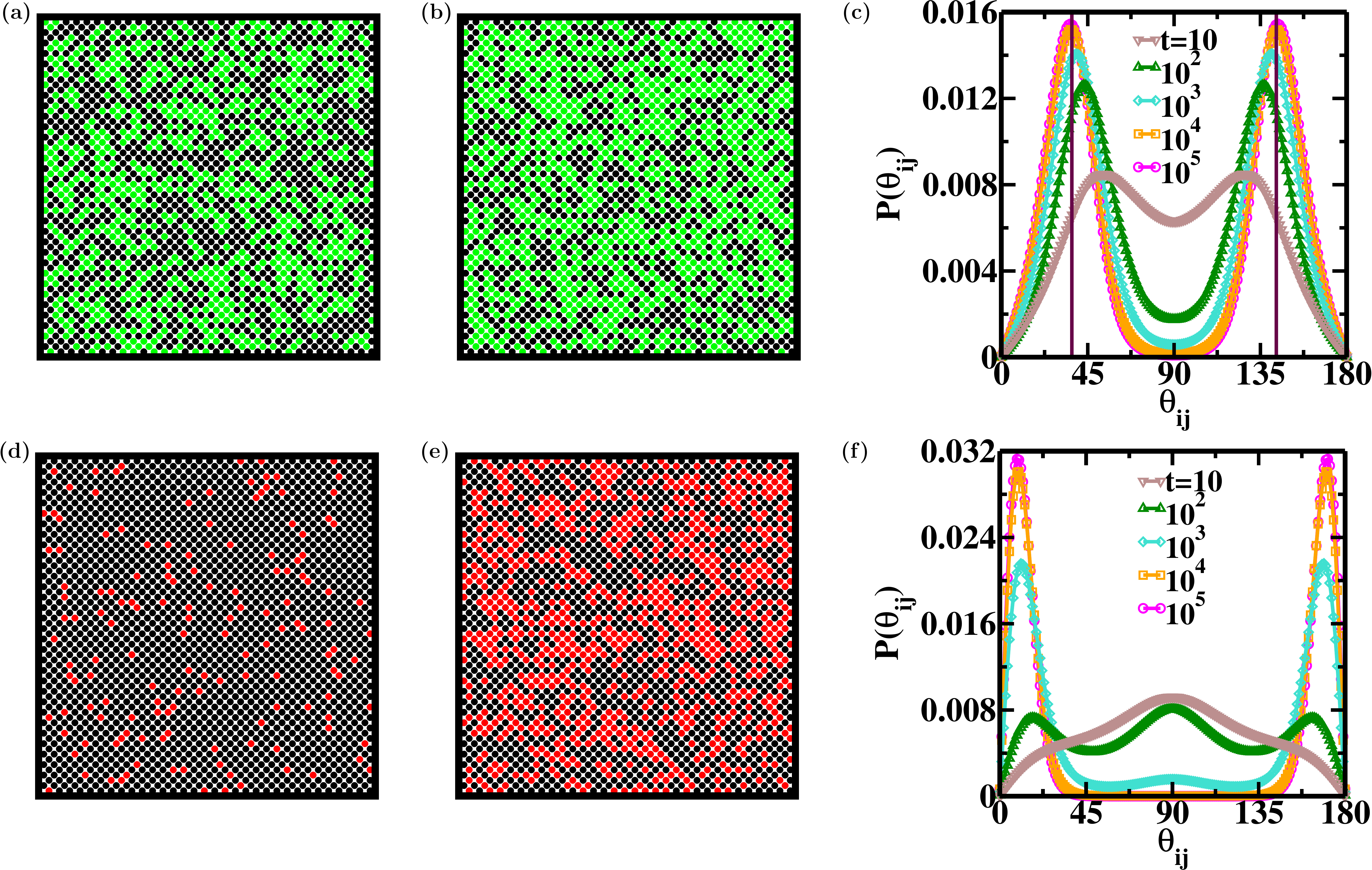}
	\caption{Horizontal ($xy$) slices (at $z=N_L/2=128$) of the bond evolution snapshots for $\lambda=-0.8$ at $t=$ (a) $10^2$, (b) $10^5$ and for $\lambda=0.8$ at $t=$ (d) $10^2$, (e) $10^5$. The interfacial defects between the nn lattice sites are coloured in black while the bonds between the nn sites (with a tolerance of $10^\circ$) are coloured in green for $\lambda=-0.8$ and in red for $\lambda=0.8$, respectively. The white regions are the corresponding lattice sites. For a better and clear view, we show only a $32^2$ corner of the entire $256^2$ slice. Plot of the nn angle distribution, $P(\theta_{ij})$ vs. $\theta_{ij}$ for (c) $\lambda=-0.8$ and (f) $\lambda=0.8$, at different times.}
	\label{figure_3}
\end{figure*}

Further insights on the coarsening morphologies are provided in Fig.~\ref{figure_3} which shows horizontal ($xy$) slices ($32^2$ corner of the entire slice taken at $z=N_L/2=128$) with bonds and interfacial defects for $\lambda=-0.8$ (top row) at $t =$ (a) $10^2$, (b) $10^5$ and for $\lambda=0.8$ (bottom row) at $t=$ (d) $10^2$, (e) $10^5$. In these slices, the orientation $\theta_{ij}$ between nn rods is evaluated. If $\theta_{ij} = \theta_{c}(\lambda)\pm 10^\circ$, the midpoint of the corresponding nn sites is coloured green (for $\lambda=-0.8$) or red (for $\lambda=0.8$). In case of a {\it defect} (i.e. $\theta_{ij} \ne \theta_{c}(\lambda)\pm 10^\circ$), the midpoint is coloured black. Morphologies in both the regimes are isotropic and coarsen with time. This is evident from the plot of nn angle distribution $P(\theta_{ij})$ vs. $\theta_{ij}$ at different times shown in Fig.~\ref{figure_3}(c) for $\lambda=-0.8$ and Fig.~\ref{figure_3}(d) for $\lambda=0.8$. As expected, the distributions for $\lambda = -0.8$ in Fig.~\ref{figure_3}(c) peak at $\theta_{ij}\equiv \theta_{c}\simeq36.7^\circ$ and $143.3^\circ$ as time evolves. For $\lambda=0.8$ in Fig.~\ref{figure_3}(f), the morphologies evolve toward the GS corresponding to $\theta_{ij}\equiv\theta_{u}=0$ (or $\pi$). Notice the secondary peak at earlier times, due to the metastablity from $\theta_{ij}=\pi/2$. The domain growth is faster for $\lambda = 0.8$, suggested from the magnitude of the peak in these distributions at later times. An important point of distinction between the canted and uniform morphologies is the distinct nature of defects - {\it interfaces} in the canted state and {\it blobs} in the uniform state. We will make quantitative statements about them shortly. 
 
\begin{figure}[h]
	\onefigure[scale=0.58]{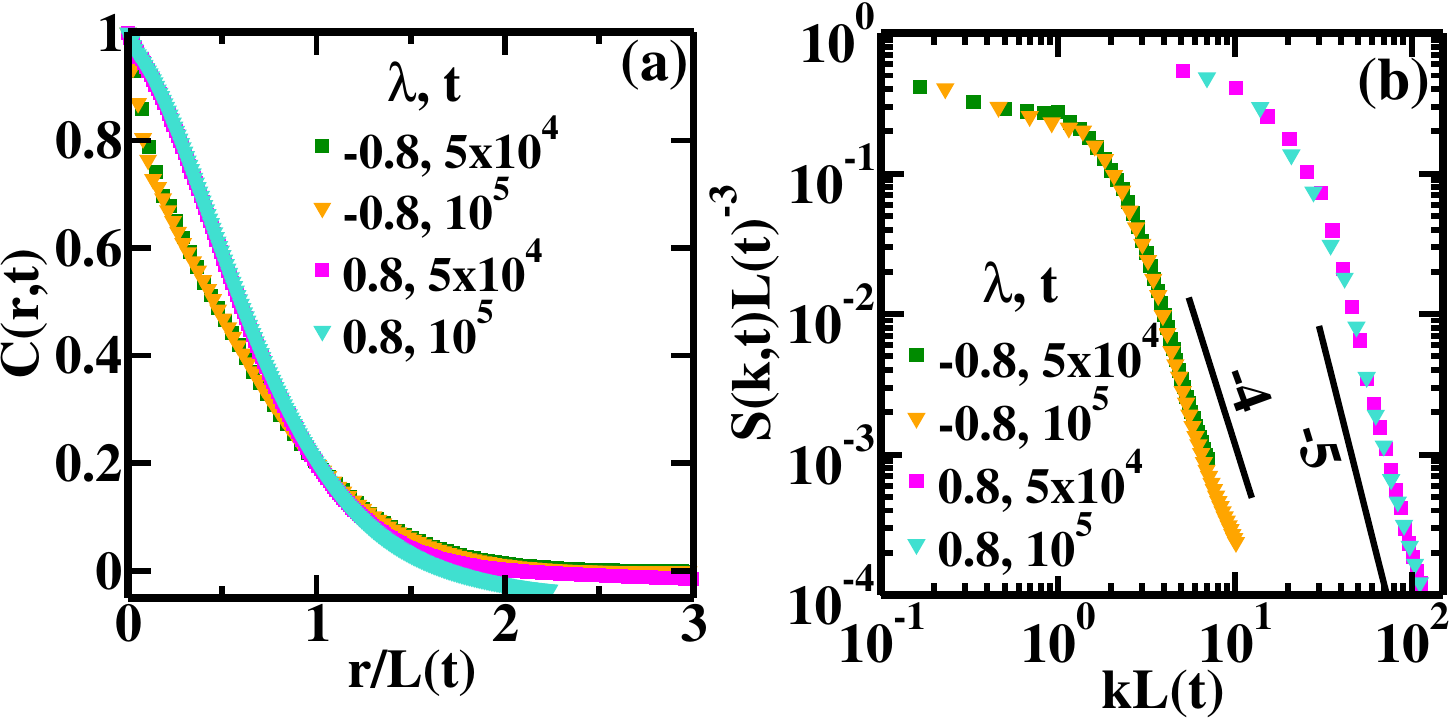}
	\caption{(a) Scaled correlation functions, $C(r,t)$ vs. $r/L(t)$, for $\lambda=-0.8$ (green coloured symbols) and $\lambda=0.8$ (red coloured sysmbols) at different times $t$ (in MCS). These plots clearly show that the GLL model exhibits dynamical scaling but not SU. (b) Scaled structure factors, $S(k,t)L(t)^{-3}$ vs. $kL(t)$ on a log-log plot, corresponding to the data sets in (a). The solid lines denote the relevant Porod tails.}
	\label{figure_4}
\end{figure}

Next, we quantify the domain morphologies. The usual probe is the correlation function: $C(\vec{r},t) = \frac{1}{N} \sum_{i}^{N} [\langle \psi(\vec{r}_i,t)\psi(\vec{r}_i + \vec{r},t) \rangle - \langle \psi(\vec{r}_i,t) \rangle \langle \psi(\vec{r}_i + \vec{r},t) \rangle]$, where $\psi(\vec{r}_i,t)$ is the order parameter, and $\langle \cdot\cdot\cdot \rangle$ represents the thermal average. Small-angle scattering experiments yield the structure factor $S(\vec{k},t) = \int\!\mathrm{d}\vec{r}~\exp(-i\vec{k}\cdot\vec{r})~C(\vec{r},t)$, where $\vec{k}$ is the wave-vector of the scattered beam. A characteristic length scale $L(t)$ is usually defined as the distance at which $C(\vec{r},t)$ decays to, say, $0.2$ times its maximum value. If the domain growth is characterized by a unique length scale $L(t)$, then $C(\vec{r},t)$ and $S(\vec{k},t)$ show the dynamical scaling property  \cite{Puri_2004,Puri_2009}: $C(\vec{r},t) = g(r/L)$; $S(\vec{k},t) =  L^d f(kL)$. {\color{black} For each value of $\lambda$, we calculate $C(\vec{r},t)$ using the local order parameter $\psi(\vec{r}_i,t) \equiv P_2(\cos \theta_i)$.} Fig.~\ref{figure_4}(a) shows the scaled correlation function, $C(r,t)$ vs. $r/L(t)$, at $t$ (in MCS) $=5\times 10^4,\ 10^5$ for $\lambda = -0.8$ and $\lambda= 0.8$. The data for different $t$-values exhibits an excellent collapse indicating that the domain growth exhibits dynamical scaling for both values of $\lambda$. However, the small-$r$ behavior for the two values of $\lambda$ is distinct. The scaling function is not robust with respect to $\lambda$ or in other words does not exhibit {\it super-universality}. 

\begin{figure}[h]
	\onefigure[scale=0.305]{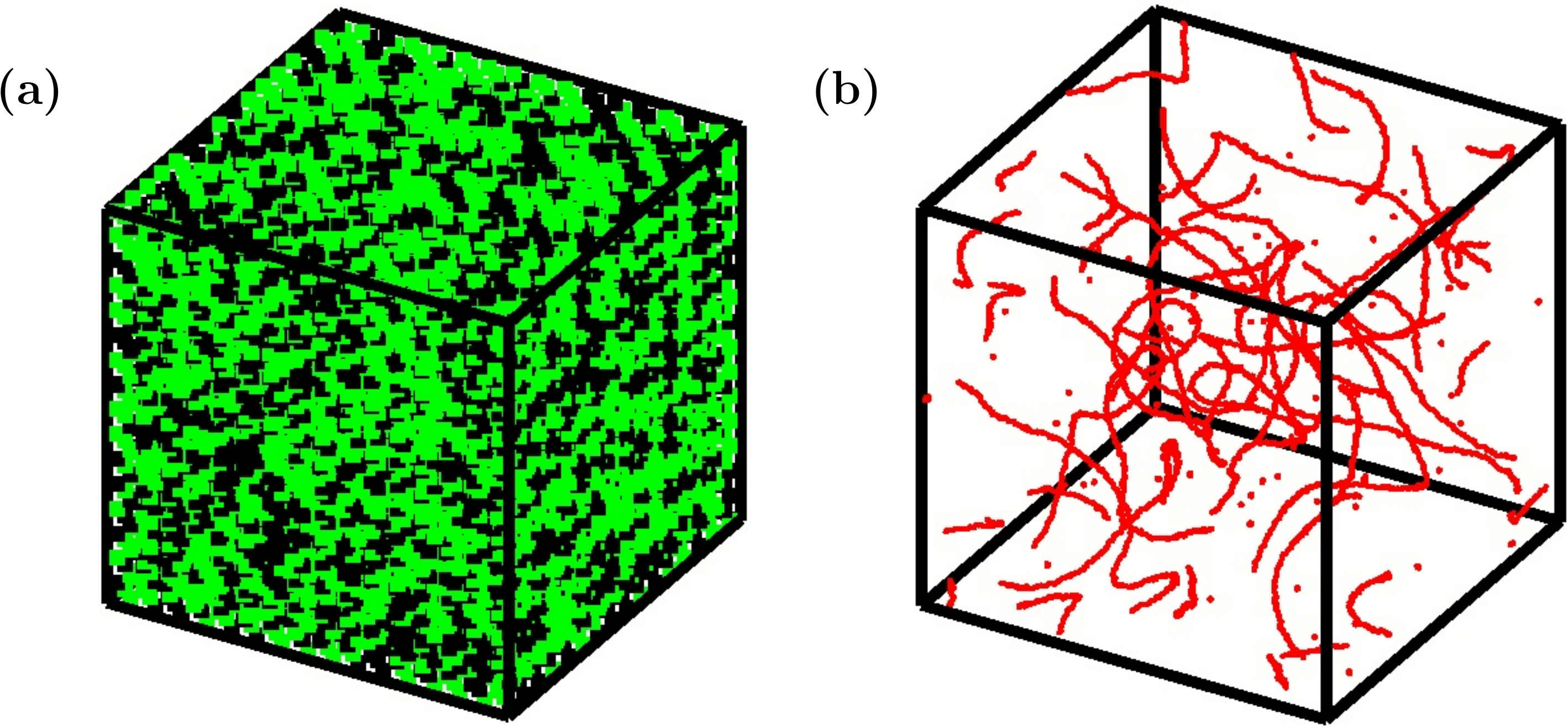}
	\caption{(a) Interfacial defects (in black colour) between canted domains (in green colour) for $\lambda=-0.8$ at $t=10^5$. We show only a $16^3$ corner of the entire lattice for a better and clear view. (b) String defects for $\lambda=0.8$ at $t=10^5$. The system size is $256^3$.}
	\label{figure_5}
\end{figure}

What are the implications of the absence of {\it super-universality} in Fig.~\ref{figure_4}(a)? Let us examine the corresponding structure factor $S(k,t)$ - especially the asymptotic (large-$k$) tail which contains information about the defects in the system. Continuous $O(n)$ spin models, {\color{black} whose Hamiltonians are invariant under rotation}, exhibit the {\it generalized Porod law} with the asymptotic form: $S(k,t) \sim k^{-(d+n)}$ {\color{black}\cite{Bray_2002,Puri_2004,Puri_2009}}. For $n=1$, the defects are interfaces, and the corresponding scattering function is the {\it Porod law}. For $n>1$, the different topological defects are vortices ($n=2,\ d=2$), strings ($n=2,\ d=3$), and monopoles or hedgehogs ($n=3,\ d=3$). Fig.~\ref{figure_4}(b) shows the plot of $L(t)^{-3}S(k,t)$ vs. $kL(t)$, corresponding to the data sets in Fig.~\ref{figure_4}(a), on a double log scale. {\color{black}The data for $\lambda=0.8$ has been shifted for clarity.} For $\lambda = -0.8$ (canted regime), the data exhibits {\it Porod decay} $S(k,t) \sim k^{-(d+1)} = k^{-4}$ for $d=3$ characteristic of scattering from sharp interfaces. This is surprising because $\mathcal{S}$ is a continuous order parameter. For $\lambda = 0.8$ (uniform regime), we find that the asymptotic tail exhibits the {\it generalized Porod decay} $S(k,t) \sim k^{-(d+2)} = k^{-5}$, indicating the predominance of string defects in the ordering kinetics. The distinct dominating defect structures in the two regimes are indeed evident in Fig.~\ref{figure_5}.

{\color{black}The defect morphology for $\lambda = -0.8$ in Fig.~\ref{figure_5}(a) shows interfaces (black colour) separating the canted domains (green colour). It is plotted the same way as the slices in Fig.~\ref{figure_3}. 
The string defects for $\lambda = 0.8$ in Fig.~\ref{figure_5}(b) have been identified by considering all the possible square plaquettes in the cubic lattice and checking whether the nematic rods in the plaquette rotate by $180^\circ$ or traverse a ``non-contractible loop'' \cite{Zapotocky_1995,Pelcovits_2001}. The observation of two distinct defect structures in the canted and uniform regimes is novel, and also the most interesting result of our study.}

\begin{figure}[h]
	\onefigure[scale=0.21]{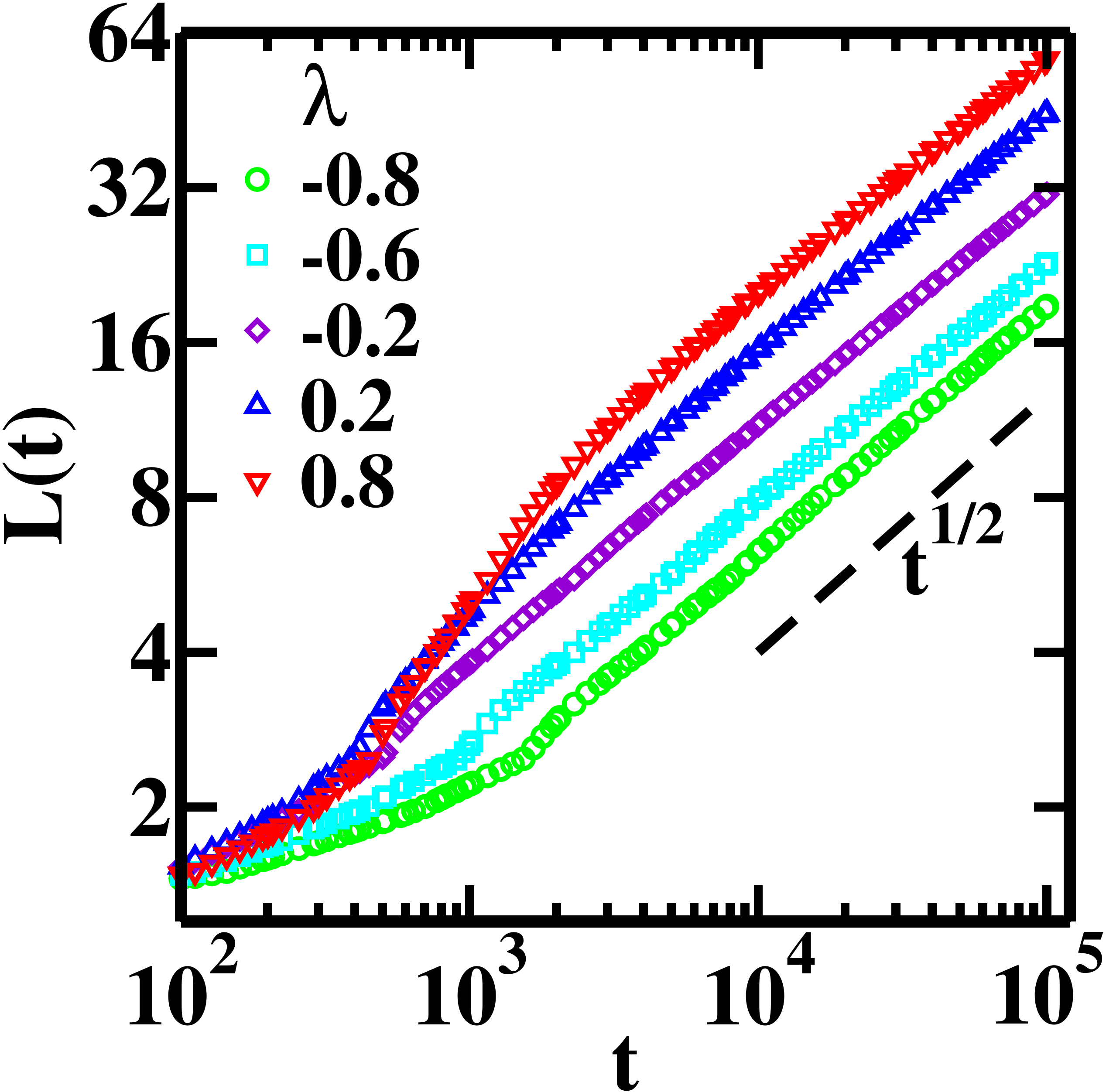}
	\caption{Plot of the characteristic length scale, $L(t)$ vs. $t$ on a log-log scale, for different values of $\lambda$ in the $d=3$ GLL model. The dashed line indicates the LAC law: $L(t) \sim t^{1/2}$.}
	\label{figure_6}	
\end{figure}

Finally, we determine the domain growth laws $[L(t)$ vs. $t]$ for the canted and uniform regimes. They reveal important details of the free-energy landscape and relaxation time scales in the system. For example, pure isotropic systems with non-conserved dynamics obey the Lifshitz-Allen-Cahn law (LAC): $L(t) \sim t^{1/2}$ \cite{Allen_1979}.  On the other hand, pure isotropic systems with conserved kinetics and diffusive transport follow the Lifshitz-Slyozov (LS) law: $L(t) \sim t^{1/3}$ \cite{Lifshitz_1961}. These growth laws are characteristic of systems with no energy barriers to coarsening and a unique relaxation timescale. Fig.~\ref{figure_6} shows $L(t)$ vs. $t$ on a log-log scale for five distinct values of $\lambda: -0.8, \ -0.6, \ -0.2, \ 0.2, \ 0.8$. The dashed line with slope 1/2 is a guide to the eye. The system obeys the LAC law, i.e., $L(t)\sim t^{1/2}$ for {\it all} values of $\lambda$, although the growth is via the annihilation of different dominating defect structures in the canted and uniform regimes. As noticed earlier in the context of Fig.~\ref{figure_2}, the domain growth is faster with increasing values of $\lambda$. The metastability in the energy for $\lambda=0.8$ is an initial barrier which has to be overcome before domain growth can proceed. 

Let us conclude this paper with a summary and discussion of our results. We have undertaken a comprehensive MC study of the kinetics of ordering in nematic liquid crystals (NLCs) after a deep thermal quench. They are modelled by the {\it generalised Lebwohl Lasher} (GLL) model ($d=3,n=3$) which incorporates the fourth order Legendre interactions $P_4(\cos\theta_{ij})$ in the Lebwohl Lasher (LL). (Here $\theta_{ij}$ is the anglular orientation between neighbouring nematogens $i$ and $j$.) Consequently, the inter-molecular potential is mimicked accurately, as evidenced from the convergence between experimental observations and corresponding numerical evaluations. 

The ground state (GS) characteristics are found to vary with parameter $\lambda$, which represents the relative strength of the $P_4$ interactions with respect to the $P_2$ interactions. For $\lambda <-0.3$, we observe a hitherto unreported {\it canted phase}, where the ground state (GS) configuration has nearest neighbour (nn) nematogens aligned at angle:
\begin{equation}
\theta_c(\lambda)=\cos^{-1}\left[\left(15\lambda-6\right)/\left(35\lambda\right)\right]^{1/2}.
\end{equation} 
For $\lambda \geq-0.3$, we observe a {\it uniform phase} where the alignment between nn rods is at an angle $\theta_u(\lambda)=0,\pi$. The coarsening morphologies exhibit {\it dynamical scaling}, implying the presence of a unique length scale for each value of $\lambda$. So the scaling function lacks {\it super-universality} as it is not robust with respect to $\lambda$. The structure factor tail in the canted regime obeys the Porod law $S(k,t)\sim k^{-4}$ characteristic of interfacial defects. This is unexpected, as the GLL model is characterised by a continuous order parameter. The uniform regime on the other hand, exhibits the expected {\it generalized Porod decay} $S(k,t)\sim k^{-5}$. The coarsening dynamics in this regime is governed by the anhilation of {\it string defects}. Finally, the domain growth obeys the Lifshitz-Allen-Cahn law $L(t)\sim t^{1/2}$ for all values of $\lambda$, but the growth is slower in the canted regime. 

Has the canted phase been observed in NLCs? The magnitudes of the second and fourth order Legendre order parameters ($\langle P_2 \rangle$ and $\langle P_4 \rangle$) in LC samples can be obtained using Raman confocal microspectrometry, but these experimental measurements are rather difficult and scarce. Fortunately for us, they have been reported in uniaxial LCs p(DR1M-\textit{co}-MMA), an azobenzene copolymer \cite{Labarthet_1998,Labarthet_2000}. The authors investigated the influence of $\langle P_4 \rangle$ relative to $\langle P_2 \rangle$, and indeed observed that the angular distribution function $[P(\theta_{ij})$ vs. $\theta_{ij}]$ exhibits a peak at angles other than 0 (or $\pi$)! However, the possible value of the model parameter $\lambda$ is not clear from these measurements.

What could be the consequences of the novel {\it canted} phase for other physical settings? The simplistic LL model has been widely used in the literature to study diverse phenomena such as ordering strategies in active matter \cite{Das_2017}, buckling and wrinkling of membranes \cite{Saikia_2017}, frustrations in orientational order due to curved geometries \cite{Mbanga_2012} and the elasticity of LC elastomers \cite{Selinger_2004}, to name a few. With the customary discrepancies between experimental observations and the corresponding numerical evaluations via the LL model, it is tempting to believe that the GLL model will offer an accurate description to capture the phenomena driven by orientational ordering. As a corollary, the consequences of the canted phase in these systems could be intriguing. {\color{black}It may also be worthwhile to expand the scope of the GLL model for the large class of colloidal systems with rod shaped molecules which yield a variety of {\it plastic} (or soft) crystals that exhibit parallel and crossed orientations  \cite{Liu_2015,Murphy_2016}. We hope that our study initiates work in these unexplored directions.}      

\acknowledgments
NB acknowledges UGC, India for a junior research fellowship. VB acknowledges DST, India for MATRICS and DST-UKIERI for research grant. NB and VB gratefully acknowledge the High Performance Computing (HPC) facility at IIT Delhi for computational resources.

\bibliographystyle{eplbib.bst}
\bibliography{Ref}

\begin{thebibliography}{10}
\expandafter\ifx\csname url\endcsname\relax\def\url#1{\texttt{#1}}\fi

\bibitem{Stephen_1974}
\Name{Stephen M.~J. \and Straley J.~P.} \REVIEW{Rev. Mod.
  Phys.}{46}{1974}{617}.

\bibitem{deGennes_1995}
\Name{de~Gennes P.~G. \and Prost J.} \Book{The Physics of Liquid Crystals}
  (Oxford: Oxford University Press) 1995.

\bibitem{Priestly_2012}
\Name{Priestly E.} \Book{Introduction to Liquid Crystals} (Springer) 2012.

\bibitem{Andrienko_2018}
\Name{Andrienko D.} \REVIEW{Journal of Molecular Liquids}{267}{2018}{520}.

\bibitem{Chen_2018}
\Name{Chen H.-W., Lee J.-H., Lin B.~Y., Chen S. \and Wu S.~T.} \REVIEW{Light:
  Sci. Appl.}{7}{2018}{17168}.

\bibitem{Chuang_1991}
\Name{Chuang I., Durrer R., Turok N. \and Yurke B.}
  \REVIEW{Science}{251}{1991}{1336}.

\bibitem{Pargellis_1991}
\Name{Pargellis A., Turok N. \and Yurke B.} \REVIEW{Phys. Rev.
  Lett.}{67}{1991}{1570}.

\bibitem{Lavrentovich_2001}
\Name{Lavrentovich O.~D., Pasini P., Zannoni C. \and Žumer S.} \Book{Defects
  in Liquid Crystals: Computer Simulations, Theory and Experiments} (Springer)
  2001.

\bibitem{Wang_2016}
\Name{Wang X., Miller D.~S., Bukusoglu E., de~Pablo J.~J. \and Abbott N.~L.}
  \REVIEW{Nature Materials}{15}{2016}{106}.

\bibitem{Kim_2018}
\Name{Kim D.~S., Copar S., Tkalec U. \and Yoon D.~K.} \REVIEW{Science
  Advances}{4}{2018}{}.

\bibitem{Sandford_2020}
\Name{Sandford~O’Neill J.~J., Salter P.~S., Booth M.~J., Elston S.~J. \and
  Morris S.~M.} \REVIEW{Nature Communications}{11}{2020}{2203}.

\bibitem{Reddy_2006}
\Name{Reddy R.~A. \and Tschierske C.} \REVIEW{J. Mater. Chem.}{16}{2006}{907}.

\bibitem{Mertelj_2013}
\Name{Mertelj A., Lisjak D., Drofenik M. \and Copic M.}
  \REVIEW{Nature}{504}{2013}{237}.

\bibitem{Maier_1958}
\Name{Maier W. \and Saupe A.} \REVIEW{Zeitschrift für Naturforschung A}{13}{01
  Jul. 1958}{564 }.

\bibitem{Saupe_1968}
\Name{Saupe A.} \REVIEW{Angewandte Chemie International Edition in
  English}{}{1968}{97}.

\bibitem{Humphries_1972}
\Name{Humphries R.~L., James P.~G. \and Luckhurst G.~R.} \REVIEW{J. Chem.
  Soc.{,} Faraday Trans. 2}{68}{1972}{1031}.

\bibitem{Sweet_1967}
\Name{Sweet J.~R. \and Steele W.~A.} \REVIEW{The Journal of Chemical
  Physics}{47}{1967}{3022}.

\bibitem{Lebwohl_1972}
\Name{Lebwohl P.~A. \and Lasher G.} \REVIEW{Phys. Rev. A}{6}{1972}{426}.

\bibitem{Zhang_1992}
\Name{Zhang Z., Mouritsen O.~G. \and Zuckermann M.~J.} \REVIEW{Phys. Rev.
  Lett.}{69}{1992}{2803}.

\bibitem{Hamley_1996}
\Name{Hamley I.~W., Garnett S., Luckhurst G.~R., Roskilly S.~J., Sedon J.~M.,
  Pedersen J.~S. \and Richardson R.~M.} \REVIEW{The Journal of Chemical
  Physics}{104}{1996}{10046}.

\bibitem{Zhang_1993_MolPhys}
\Name{Zhang Z., Zuckermann M.~J. \and Mouritsen O.~G.} \REVIEW{Molecular
  Physics}{80}{1993}{1195}.

\bibitem{Chiccoli_1997}
\Name{Chiccoli C., Pasini P. \and Zannoni C.} \REVIEW{International Journal of
  Modern Physics B}{11}{1997}{1937}.

\bibitem{Fuller_1985}
\Name{Fuller G., Luckhurst G. \and Zannoni C.} \REVIEW{Chemical
  Physics}{92}{1985}{105 }.

\bibitem{Pelcovits_2001}
\Name{Priezjev N.~V. \and Pelcovits R.~A.} \REVIEW{Phys. Rev.
  E}{64}{2001}{031710}.

\bibitem{Bray_2002}
\Name{Bray A.~J.} \REVIEW{Advances in Physics}{51}{2002}{481}.

\bibitem{Puri_2004}
\Name{Puri S.} \REVIEW{Phase Transitions}{77}{2004}{407}.

\bibitem{Puri_2009}
\Name{Puri S. \and Wadhawan V.} \Book{Kinetics of phase transitions} (CRC
  Press, Boca Raton) 2009.

\bibitem{Zapotocky_1995}
\Name{Zapotocky M., Goldbart P.~M. \and Goldenfeld N.} \REVIEW{Phys. Rev.
  E}{51}{1995}{1216}.

\bibitem{Yeomans_2001}
\Name{Denniston C., Orlandini E. \and Yeomans J.~M.} \REVIEW{Phys. Rev.
  E}{64}{2001}{021701}.

\bibitem{Bhattacharjee_2008}
\Name{Bhattacharjee A.~K., Menon G.~I. \and Adhikari R.} \REVIEW{Phys. Rev.
  E}{78}{2008}{026707}.

\bibitem{Blundell_1992}
\Name{Blundell R.~E. \and Bray A.~J.} \REVIEW{Phys. Rev. A}{46}{1992}{R6154}.

\bibitem{Toyoki_1993}
\Name{Toyoki H.} \REVIEW{Phys. Rev. E}{47}{1993}{2558}.

\bibitem{Dutta_2005}
\Name{Dutta S. \and Roy S.~K.} \REVIEW{Phys. Rev. E}{71}{2005}{026119}.

\bibitem{Singh_2012}
\Name{Singh A., Ahmad S., Puri S. \and Singh S.} \REVIEW{{EPL} (Europhysics
  Letters)}{100}{2012}{36004}.

\bibitem{Singh_2014}
\Name{Singh A., Ahmad S., Puri S. \and Singh S.} \REVIEW{Eur. Phys. J.
  E}{37}{2014}{2}.

\bibitem{Singh_2013}
\Name{Singh A. \and Singh S.} \REVIEW{Eur. Phys. J. E}{36}{2013}{122}.

\bibitem{Bray_1993}
\Name{Bray A.~J., Puri S., Blundell R.~E. \and Somoza A.~M.} \REVIEW{Phys. Rev.
  E}{47}{1993}{R2261}.

\bibitem{Wickam_1997}
\Name{Wickham R.~A.} \REVIEW{Phys. Rev. E}{56}{1997}{6843}.

\bibitem{Zhang_1993_PRE}
\Name{Zhang Z., Mouritsen O.~G. \and Zuckermann M.~J.} \REVIEW{Phys. Rev.
  E}{48}{1993}{2842}.

\bibitem{Bradac_2011}
\Name{Bradac Z., Kralj S. \and Zumer S.} \REVIEW{The Journal of Chemical
  Physics}{135}{2011}{024506}.

\bibitem{Billeter_1999}
\Name{Billeter J.~L., Smondyrev A.~M., Loriot G.~B. \and Pelcovits R.~A.}
  \REVIEW{Phys. Rev. E}{60}{1999}{6831}.

\bibitem{Care_2005}
\Name{Care C.~M. \and Cleaver D.~J.} \REVIEW{Reports on Progress in
  Physics}{68}{2005}{2665}.

\bibitem{Landau_2005}
\Name{Landau D.~P. \and Binder K.} \Book{A Guide to Monte Carlo Simulations in
  Statistical Physics} (Cambridge: Cambridge University Press) 2005.

\bibitem{Parker_1997}
\Name{Neal M.~P., Parker A.~J. \and Care C.~M.} \REVIEW{Molecular
  Physics}{91}{1997}{603}.

\bibitem{Withers_2002}
\Name{Withers I.~M., Care C.~M., Neal M.~P. \and Cleaver D.~J.}
  \REVIEW{Molecular Physics}{100}{2002}{1911}.

\bibitem{Allen_1979}
\Name{Allen S.~M. \and Cahn J.~W.} \REVIEW{Acta Metallurgica}{27}{1979}{1085}.

\bibitem{Lifshitz_1961}
\Name{Lifshitz I. \and Slyozov V.} \REVIEW{Journal of Physics and Chemistry of
  Solids}{19}{1961}{35}.

\bibitem{Labarthet_1998}
\Name{Lagugne~Labarthet F., Buffeteau T. \and Sourisseau C.} \REVIEW{The
  Journal of Physical Chemistry B}{102}{1998}{5754}.

\bibitem{Labarthet_2000}
\Name{Labarthet F.~L., Buffeteau T. \and Sourisseau C.} \REVIEW{Applied
  Spectroscopy}{54}{2000}{699}.

\bibitem{Das_2017}
\Name{Das R., Kumar M. \and Mishra S.} \REVIEW{Scientific
  Reports}{7}{2017}{7080}.

\bibitem{Saikia_2017}
\Name{Saikia L., Sarkar T., Thomas M., Raghunathan V.~A., Sain A. \and Sharma
  P.} \REVIEW{Nature Communications}{8}{2017}{1160}.

\bibitem{Mbanga_2012}
\Name{Mbanga B.~L., Grason G.~M. \and Santangelo C.~D.} \REVIEW{Phys. Rev.
  Lett.}{108}{2012}{017801}.

\bibitem{Selinger_2004}
\Name{Selinger J.~V. \and Ratna B.~R.} \REVIEW{Phys. Rev. E}{70}{2004}{041707}.

\bibitem{Liu_2015}
\Name{Liu B., Besseling T.~H., van Blaaderen A. \and Imhof A.} \REVIEW{Phys.
  Rev. Lett.}{115}{2015}{078301}.

\bibitem{Murphy_2016}
\Name{Murphy R.~P., Hong K. \and Wagner N.~J.}
  \REVIEW{Langmuir}{32}{2016}{8424}.

\end{thebibliography}

\end{document}